\documentstyle[editedvolume,psfig]{crckapb} 

\begin{opening}
\title{FIRST DREDGE-UP AND FURTHER MIXING MECHANISMS \protect\\
 ALONG THE RED GIANT BRANCH IN FIELD STARS}

\author{E. CARRETTA}
\author{R.G. GRATTON}
\institute{Osservatorio Astronomico di Padova\\
           (Italy), e-mail: carretta@pd.astro.it, gratton@pd.astro.it}
\author{C. SNEDEN}
\institute{Dept. of Astronomy, Univ. of Texas at Austin, TX\\
           (USA), e-mail: chris@verdi.as.utexas.edu}
\author{A. BRAGAGLIA}
\institute{Osservatorio Astronomico di Bologna\\
           (Italy), e-mail: angela@bo.astro.it}

\end{opening}

\begin{document}


\section{Mixing along the Red Giant Branch (RGB)}

As a small mass star evolves up the RGB, the outer convective envelope expands
inward and penetrates into the CN-cycle processed interior regions  
({\bf first dredge-up}).
Some further mixing is possible in the latest phases of the RGB (see
Charbonnel 1995: C95), when the advancing H-burning shell reaches the
chemical discontinuity left behind by the receding convective envelope.
Thereafter, no mean molecular weight gradient is present between the
convective envelope and the near vicinity of the shell, and is possible that
circulation currents give rise to further mixing.

Here we test this scenario presenting abundances for Fe, Li, C, N, O for about
60 metal-poor ($-2<$[Fe/H]$<-1$) field stars in different evolutionary stages.
Details of the analysis will be given in a forthcoming paper.
The main results of our study are presented in Figure 1, where abundances as a
function of luminosity are shown.

There are two distinct mixing/dilution episodes along the RGB evolution
of small mass  field stars:

1) the {\bf first dredge-up} follows canonical predictions (see e.g.       
C95). This is quite clear in the pattern shown by elements and isotopic ratios
in panels a,b,c. The luminosity is about $\log{L/L_\odot}=1$; Li is
diluted by about a factor of 20, $^{12}$C abundance decreases by $\sim 0.1$~dex.

2) a {\bf second mixing episode} occurs when the star becomes brighter than the
RGB bump (and the molecular weight barrier is canceled) at
$\log{L/L_\odot}\sim 2$, again in agreement with predictions by C95. Most of
the remaining Li is destroyed, $^{12}$C abundance further decreases by a
factor $\sim 2$, and $^{12}$C/$^{13}$C ratio raises to $\sim 6$. These
values are observed also in the following evolutionary phase (RHB and
early AGB)

O and Na abundances in bright RGB and HB field stars are similar to those
observed in unevolved stars. {\bf Field stars do not display any signature} of
the Na-O anticorrelation seen amongst globular cluster giants (see e.g. Kraft
1994).
   
\begin{figure}
\psfig{figure=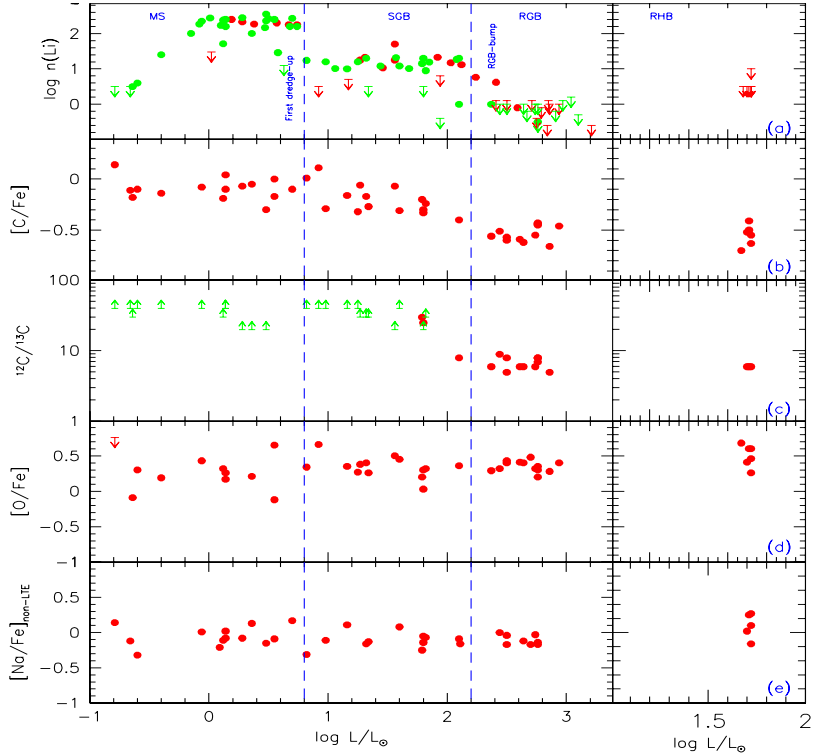,width=12.0cm,height= 10.25cm}
\caption[]{Abundances of light elements as a function of luminosity in field
stars.}\end{figure}


\end{document}